\documentclass[twocolumn,showpacs,preprintnumbers]{revtex4}
\usepackage{graphicx}
\usepackage{dcolumn}
\usepackage{bm}
\usepackage{epsfig}
\usepackage{amsmath,amssymb}
\usepackage{ulem}
\usepackage{CJKutf8}
\draft

\begin{document}

\begin{CJK}{UTF8}{<font>}
\title{ Two-proton radioactivity of exotic nuclei beyond proton drip-line }
\author{Yanzhao Wang$^{1,2,3}$}
\email{yanzhaowang09@126.com}
\author{Jianpo Cui$^{1,2}$}
\author{Yonghao Gao$^{1,2}$}
\author{Jianzhong Gu$^{3}$}
\email{jzgu1963@ciae.ac.cn}
\affiliation{$^1$ Department of Mathematics and Physics, Shijiazhuang Tiedao University,
Shijiazhuang 050043, China\\
$^2$ Institute of Applied Physics, Shijiazhuang Tiedao University,
Shijiazhuang 050043, China \\
$^3$ China Institute of Atomic Energy, P. O. Box 275 (10), Beijing 102413,
China }
\date{\today }

\begin{abstract}
To search for new candidates of the true and simultaneous two-proton ($2p$) radioactivity, the $2p$ decay energies (\textit{Q}$_{2p}$) are extracted by the Weizs\"{a}cker-Skyrme-4 (WS4) model, the finite-range droplet model (FRDM), the Kourra-Tachibaba-Uno-Yamada (KTUY) model and the Hartree-Fock-Bogoliubov mean-field model with the BSk29 Skyrme interaction (HFB29). Then, the $2p$ radioactivity half-lives are calculated within the generalized liquid drop model (GLDM) by inputting the four types of \textit{Q}$_{2p}$ values. By the energy and half-life constraints, it is found that the probable $2p$ decay candidates are the nuclei beyond the proton-drip line in the region of \textit{Z} $<$50 or \textit{Z}$\leq 50$ based on each nuclear mass model. In the region beyond \textit{Z}=50, the $2p$-decaying candidates are predicted only using the HFB29 mass model.
Finally, the competition between the true $2p$ radioactivity and $\alpha$-decay for the nuclei above the \textit{N}=\textit{Z}=50 shell closures is discussed. It is shown that $^{101}$Te, $^{111}$Ba and $^{114}$Ce prefer to $2p$ radioactivity and the dominant decay mode of $^{107}$Xe and $^{116}$Ce is $\alpha$-decay.
\end{abstract}
\maketitle

\section{Introduction} \label{sec1}

With the construction of a new generation radioactive beam facilities and the development of the new detection technology, study on the exotic decay properties of unstable nuclei has been a hot subject in nuclear physics~\cite{Faestermann2013,Pfutzner2012,Blank2008a,Blank2008b,Sonzogni2002}. Nowadays, the proton radioactivity has been recognized as one of the exotic decay modes and paid attention by many researchers~\cite{Pfutzner2012,Blank2008a,Blank2008b,Sonzogni2002}. At the beginning of the 1960s, the proton radioactivity was proposed in very proton-rich nuclei by Goldansky, Zel'dovich and Karnaukhov~\cite{Goldansky1960,Zel1960,Kar1961}.
In the region near the proton drip line, the one-proton ($1p$) and $2p$ radioactivity were observed in recent decades~\cite{Pfutzner2012,Blank2008a,Blank2008b,Sonzogni2002}.
For the $1p$ radioactivity, it was first discovered from the isomer state of $^{53}$Co ($^{53}$Co$^{m}$)~\cite{Jackson1970,Cerny1970}, but the ground state $1p$ radioactivity was first observed in the decay of $^{151}$Lu and $^{147}$Tm~\cite{Hofmann1982,Klepper1982}. So far, about 25 ground state $1p$ emitters have been identified~\cite{Sonzogni2002} and the $1p$ radioactivity has become a powerful tool to extract the nuclear structure details, such as the sequences of single-particle energies, the wave function of an emitted proton, nuclear masses, and deformations~\cite{Kruppa2004,Ferreira2007}.
For the $2p$ radioactivity, it was another new and exotic decay mode with the simultaneous emission of the two protons from the unbound even-Z nuclei because of the pairing effect. In 1960, Zel'dovich predicted the possibility that a pair of protons might emit from a nucleus~\cite{Zel1960} and it was defined as \textquotedblleft $2p$ radioactivity\textquotedblright by Goldansky~\cite{Goldansky1960,Goldansky1961}. More than 40 years later, the ground-state true $2p$ radioactivity (\textit{Q}$_{2p}>0$ and \textit{Q}$_{p}<0$, where \textit{Q}$_{p}$ is the released energy of the $1p$ radioactivity.)~\cite{Pfutzner2012} was firstly observed from $^{45}$Fe at GSI~\cite{Pfutzner2002} and at GANIL~\cite{Giovinazzo2002}, respectively. Then, the true $2p$ radioactivity was observed from the ground states of $^{54}$Zn~\cite{Blank2005}, $^{48}$Ni~\cite{Dossat2005}, $^{19}$Mg~\cite{Mukha2007} and $^{67}$Kr~\cite{Goigoux2016}. In addition to the ground-state true $2p$ radioactivity, in fact, before 2000 ones observed the $2p$ radioactivity from the very short-lived nuclear ground states and excited states. For the $2p$ radioactivity with a very short half-life, it was discovered from $^{6}$Be~\cite{Whaling1966}, $^{12}$O~\cite{KeKelis1978} and $^{16}$Ne~\cite{KeKelis1978}, whose decaying states have a large width so that the $2p$ emitter states and the $1p$ daughter states overlap with each other. This case was interpreted as the \textquotedblleft democratic decay\textquotedblright in later studies~\cite{Pfutzner2012,Blank2008a,Blank2008b,Bochkarev1989}. For the $2p$ radioactivity from the excited states, it includes the $\beta $-delayed two-proton emission~\cite{Cable1983} and the two-proton emission from excited states populated in nuclear reactions~\cite{Bain1996}. Nowadays, the above mentioned three types of $2p$ radioactivity have been important frontiers in the study of nuclear radioactivity.

To describe the $2p$ radioactivity, many approaches have been proposed which can be roughly divided into two completely different cases.
One case considers the two emitted protons from a parent nucleus are correlated strongly due to the proton-proton attraction. This decay process is called
as \textquotedblleft $^{2}$He\textquotedblright cluster emission~\cite{Janecke1965,Brown1991,Naza1996,Ormand1997,Barker2001,Brown2003,Grigorenko2007a,Delion2013,Rotureau2006,Gonalves2017,cui20,Olsen2013}.
The other one refers to the completely uncorrelated emission, which is usually named as a three-body radioactivity~\cite{Olsen2013,Galitsky1964,Danilin1993,Grigorenko2000,Vasilevsky2001,Grigorenko2002,Grigor2003,Descouvemont2006,Grig2007,Garrido2008,Alvarez2008b,Alvarez2010,Blank2011}.
The experimental half-lives of the $2p$ radioactivity are reproduced more or less satisfactorily by these models.

On the other hand, many candidates of the $2p$ radioactivity were predicted by various models~\cite{Brown1991,Gonalves2017,cui20,Olsen2013,Saxena2017,Tavares2018,Sreeja2019,Singh2012}. As a matter of fact, the earliest determination for the candidates of the $2p$ radioactivity can date back to the pioneering work of Goldansky~\cite{Goldansky1960}. For the ground state $2p$ radioactivity, the predicted candidates are the nuclei with \textit{Z}$<$38. So, it is interesting to know whether or not the $2p$ radioactivity exists in heavier systems. Recently, Olsen \textit{et al.} delineated the full landscape of $2p$ radioactivity by the energy density functional theory with several Skyrme interactions to search for the new $2p$-emitters of the heavier nuclides with \textit{Z}$>$38~\cite{Olsen2013}. In addition, the competition between
$2p$ emission and $\alpha$-decay has been predicted for some cases with a chance of being observed (nuclei around $^{103}$Te--$^{110}$Ba)~\cite{Olsen2013}. This study shows that only in two mass regions the $2p$-decay mode is predicted to occur and close enough to be addressed by today's experiments. One region ranges from germanium to krypton and the second one is located just above tin~\cite{Olsen2013}.

However, the $2p$ radioactivity half-life is dependent strongly on the \textit{Q}$_{2p}$ value. It is well known that the \textit{Q}$_{2p}$ value can be extracted by the following expression,
\begin{equation}
\mathit{Q}_{2p}(N,Z)\approx -\mathit{S}_{2p}(N,Z)=B(N,Z-2)-B(N,Z),  \label{Q formula}
\end{equation}
where $B(N,Z)$ is the binding energy of the nuclide related to its mass $M(N,Z)$: $M(N,Z)=ZM_{H}+Nm_{n}-B(N,Z)$ (\textit{M}$_{H}$ and \textit{m}$_{n}$ are the masses of the hydrogen atom and the neutron, respectively). So, it is very important to get accurate \textit{Q}$_{2p}$ values by the nuclear mass models with high precision and strong predictive ability.

In recent years, various nuclear mass models have been developed by phenomenological and microscopic approaches~\cite{romass,liquid,WS3,WN14,FRDM,SNMF,BW2,SHFB,GHFB,RMF,ETFSI,ETF2,geng,qu,KTUY05,DZ,DZQi,INMM}. Relevant studies suggested that the WS4
and FRDM mass models have high accuracy and strong predicted ability and their accuracy is higher than that of the Hartree-Fock-Bogoliubov model with Skyrme or Gogny interactions~\cite{WN14,wang2015,cui2018}.
Nowadays, the observed $2p$ radioactivity is the true and simultaneous emission. The nucleon wave functions and nucleon-nucleon interaction are involved in the
angular momentum and correlation between the two emitted protons. Therefore, this kind of $2p$ radioactivity is more significant for the study of the exotic nuclear structure.
Thus, to search for the new $2p$-emitters of the true and simultaneous case, in this article the WS4~\cite{WN14}
and FRDM~\cite{FRDM} models will be used to extract the \textit{Q}$_{2p}$ values. In calculations, to observe the model dependence of \textit{Q}$_{2p}$ values, the HFB29~\cite{SHFB} and KTUY~\cite{KTUY05} models will also be applied.
In addition, for future measurements of the $2p$ radioactivity, based on our recent work on the $2p$ radioactivity~\cite{cui20}, the half-lives of the $2p$ radioactivity will be calculated within the successful GLDM by inputting different kinds of \textit{Q}$_{2p}$ values. At last, the competition between $2p$ radioactivity and $\alpha$-decay will be studied in the framework of the GLDM. This article is organized as follows. In Sec. II, the theoretical framework is introduced. In Sec. III, the calculated results are shown and relevant discussions are performed. In the last section, some conclusions are drawn.

\section{Theoretical framework} \label{sec2}

\subsection{Methods of extracting $Q_{2p }$ values}

In the past a few years, many nuclear mass models, such as the macroscopic-microscopic
models~\cite{romass,liquid,WS3,WN14,FRDM,SNMF,BW2}, the microscopic models
based on the mean-field concept~\cite{SHFB,GHFB,RMF,ETFSI,ETF2,geng,qu} and
other kind of models~\cite{KTUY05,DZ,DZQi,INMM}, have been developed with
rms deviations from several hundred keV to a few MeV with respect to all
known nuclear masses. Within the nuclear mass tables combining Eq.(1), the
\textit{Q}$_{2p}$ values can be obtained. In the \textit{Q}$_{2p}$ calculations, the
WS4~\cite{WN14}, FRDM~\cite{FRDM}, HFB29~\cite{SHFB} and KTUY~\cite{KTUY05} mass tables are used.

\subsection{GLDM}
In the GLDM, the process of the shape evolution from one body to two separated fragments can be described in a unified way. Its details can be found in Refs.~\cite{cui2016,GLDM,wang2010,wang2014}.

For the $2p$ radioactivity in the framework of the GLDM, we consider it as a $^{2}$He cluster emission.
It is assumed that a preformed $2p$ pair penetrates the Coulomb barrier of the parent nucleus and decays outside the barrier.
The $2p$ pair preformed nearby the surface of the parent nucleus has zero binding energy and the two protons separate quickly due to the dominance of the Coulomb repulsion after they escape from the parent nucleus. Meanwhile, the orbital angular momentum carried by the $2p$ pair satisfies the spin-parity selection rule.

In the GLDM, the $2p$ radioactivity half-life is defined as
\begin{equation}  \label{GLDM2}
T_{1/2}^{2p}=\frac{\ln 2}{\lambda_{2p}},
\end{equation}
where $\lambda_{2p}$ is the decay constant, which is expressed as
\begin{equation}  \label{decay constant1}
\lambda_{2p}=S_{2p}\nu _{0}P.
\end{equation}
Here $S_{2p}$ denotes the spectroscopic factor of the $2p$ radioactivity. It can be estimated in the cluster overlap approximation~\cite{Brown1991}, $S_{2p}=G^{2}[A/(A-2)]^{2n}\chi ^{2}$. Here, $G^{2}=(2n)!/[2^{2n}(n!)^{2}]$~\cite{Anyas1974}, and $n$ is the average principal proton oscillator quantum number given by $n$$\approx (3Z)^{1/3}-1$~\cite{Bohr1969}. $A$ and $Z$ are the mass number and the charge number of the parent nucleus, respectively. $\chi ^{2}=0.0143$, which is determined by fitting the experimental half-lives of $^{19}$Mg, $^{45}$Fe, $^{48}$Ni, and $^{54}$Zn~\cite{cui20}.
$\nu _{0}$ is the assault frequency of the $2p$ pair on the barrier of the parent nucleus and estimated by the classical method
\begin{equation}  \label{frequency}
\mathit{\nu }_{0}=\frac{1}{2R}\sqrt{\frac{2E_{2p}}{M_{2p}}},
\end{equation}
where $R$ is the radius of the parent nucleus. $E_{2p}$ and $M_{2p}$ represent
the kinetic energy and the mass of the emitted $2p$ pair, respectively.

The penetrability factor $P$ is calculated by the WKB approximation, which is expressed as
\begin{equation}  \label{penetrability0}
P=\exp \left[ -\frac{2}{\hbar }\int_{R_{\text{in}}}^{R_{\text{out}}}%
\sqrt{2B(r)[E(r)-E_{sph}]}\text{d}r\right],\
\end{equation}
where \textit{R}$_{\text{in}}$ and \textit{R}$_{\text{out}}$ are the two turning points of the WKB action integral.
Here, an approximation is used $B(r)=\mu$, and $\mu$ stands for the reduced mass of the
$2p$ pair and the residual daughter nucleus.
The macroscopic energy \textit{E}(\textit{r}) is written as
\begin{equation}  \label{etot}
E(r)=E_{V}+E_{S}+E_{C}+E_{Prox}+E_{cen},
\end{equation}
which contains the volume, surface, Coulomb, proximity, and centrifugal potential energies.

\section{Results and discussions} \label{sec3}

\begin{table*}[t]
\begin{ruledtabular}\caption{The comparison between the experimental \textit{Q}$_{2p}$ values and those extracted from the WS4~\cite{WN14}, FRDM~\cite{FRDM}, KTUY~\cite{KTUY05}, and HFB29~\cite{SHFB} nuclear mass models. All the \textit{Q}$_{2p}$ values are measured in MeV.}
\begin{tabular}{llllccc}
Nuclei & \textit{Q}$_{2p}^{\text{exp.}}$(MeV)   & \textit{Q}$_{2p}^{\text{WS4}}$(MeV) & \textit{Q}$_{2p}^{\text{FRDM}}$(MeV) &\textit{Q}$_{2p}^{\text{KTUY}}$(MeV) &\textit{Q}$_{2p}^{\text{HFB29}}$(MeV)\\
\hline
$^{19}_{12}$Mg&0.750~\cite{Mukha2007}      & --  & -- & 1.14 & --\\
$^{45}_{26}$Fe&1.100~\cite{Pfutzner2002}   &2.06 &1.89&1.19& 1.92\\
         &1.140~\cite{Giovinazzo2002}      &     &    &    &     \\
         &1.154~\cite{Dossat2005}          &     &    &    &     \\
         &1.210~\cite{Audirac2012}         &     &    &    &     \\
$^{48}_{28}$Ni&1.350~\cite{Dossat2005}     &2.54 &3.30&1.95& 3.63\\
              &1.290~\cite{Pomorski2014}   &     &    &    &     \\
              &1.310~\cite{Wang2017}       &     &    &    &     \\
$^{54}_{30}$Zn&1.480~\cite{Blank2005}      &1.98 &2.77&1.65& 1.61\\
         & 1.280~\cite{Ascher2011}         &     &    &    &     \\
$^{67}_{36}$Kr&1.690~\cite{Goigoux2016}    &3.06 &1.33&1.52& 1.94\\
\end{tabular}
\end{ruledtabular}
\end{table*}

\begin{table*}[t]
\begin{ruledtabular}\caption{The \textit{Q}$_{2p}$ values and $2p$ decay half-lives of the true and simultaneous emission for the nuclei beyond the proton-drip line. The log$_{10}T_{1/2}^{\text{2p}}$ values are measured in seconds.}
\begin{tabular}{c c c c c c c c c c c c c c c c c c c c c}
Nuclei & Mass models & \textit{Q}$_{2p}$ (MeV) &  log$_{10}T_{1/2}^{\text{2p}}$ (s)& Nuclei & Mass models & \textit{Q}$_{2p}$ (MeV) &  log$_{10}T_{1/2}^{\text{2p}}$ (s)&\\
\hline
$^{31}_{18}$Ar	&	WS4	&	1.08 	&	-9.54 	&	$^{41}_{24}$Cr	&	KTUY	&	2.13 	&	-11.10 	&	\\
$^{39}_{22}$Ti	&	WS4	&	0.98 	&	-4.62 	&	$^{58}_{32}$Ge	&	KTUY	&	2.45 	&	-8.80 	&	\\
$^{42}_{24}$Cr	&	WS4	&	1.40 	&	-6.95 	&	$^{62}_{34}$Se	&	KTUY	&	3.27 	&	-10.90 	&	\\
$^{49}_{28}$Ni	&	WS4	&	1.13 	&	-0.75 	&	$^{63}_{34}$Se	&	KTUY	&	1.78 	&	-3.45 	&	\\
$^{58}_{32}$Ge	&	WS4	&	2.68 	&	-9.80 	&	$^{65}_{36}$Kr	&	KTUY	&	2.83 	&	-8.34 	&	\\
$^{59}_{32}$Ge	&	WS4 &	1.43 	&	-1.73 	&	$^{70}_{38}$Sr	&	KTUY	&	2.80 	&	-7.18 	&	\\
$^{62}_{34}$Se	&	WS4	&	3.64 	&	-12.00 	&	$^{74}_{40}$Zr	&	KTUY	&	2.65 	&	-5.39 	&	\\
$^{63}_{34}$Se	&	WS4	&	2.39 	&	-7.38 	&	$^{78}_{42}$Mo	&	KTUY	&	3.07 	&	-6.08 	&	\\
$^{68}_{36}$Kr	&	WS4	&	1.80 	&	-2.23 	&	$^{79}_{42}$Mo	&	KTUY	&	2.19 	&	-1.07 	&	\\
$^{71}_{38}$Sr	&	WS4	&	2.92 	&	-7.72 	&	$^{86}_{46}$Pd	&	KTUY	&	3.34 	&	-5.35 	&	\\
$^{72}_{38}$Sr	&	WS4	&	1.65 	&	0.59 	&	$^{90}_{48}$Cd	&	KTUY	&	3.39 	&	-4.64 	&	\\
$^{75}_{40}$Zr	&	WS4	&	2.80 	&	-6.15 	&	$^{94}_{48}$Cd	&	KTUY	&	0.34 	&	0.00 	&	\\
$^{76}_{40}$Zr	&	WS4	&	1.85 	&	0.10 	&	$^{94}_{50}$Sn	&	KTUY	&	3.34 	&	-3.39 	&	\\
$^{79}_{42}$Mo	&	WS4	&	3.07 	&	-6.09 	&	$^{95}_{50}$Sn	&	KTUY	&	2.48 	&	1.53 	&	\\
$^{80}_{42}$Mo	&	WS4	&	2.34 	&	-2.13 	&	$^{31}_{18}$Ar	&	HFB29	&	0.43 	&	1.14 	&	\\
$^{83}_{44}$Ru	&	WS4	&	3.20 	&	-5.69 	&	$^{49}_{28}$Ni	&	HFB29	&	1.88 	&	-7.33 	&	\\
$^{84}_{44}$Ru	&	WS4	&	2.20 	&	0.08 	&	$^{58}_{32}$Ge	&	HFB29	&	1.77 	&	-4.76 	&	\\
$^{86}_{46}$Pd	&	WS4	&	3.83 	&	-7.23 	&	$^{62}_{34}$Se	&	HFB29	&	3.04 	&	-10.20 	&	\\
$^{87}_{46}$Pd	&	WS4	&	3.04 	&	-3.98 	&	$^{63}_{34}$Se	&	HFB29	&	1.80 	&	-3.61 	&	\\
$^{90}_{48}$Cd	&	WS4	&	3.90 	&	-6.65 	&	$^{65}_{36}$Kr	&	HFB29	&	3.11 	&	-9.45 	&	\\
$^{91}_{48}$Cd	&	WS4	&	3.08 	&	-3.19 	&	$^{70}_{38}$Sr	&	HFB29	&	3.38 	&	-9.47 	&	\\
$^{96}_{50}$Sn	&	WS4	&	2.67 	&	0.20 	&	$^{71}_{38}$Sr	&	HFB29	&	2.15 	&	-3.56 	&	\\
$^{30}_{18}$Ar	&	FRDM	&	1.22 	&	-10.65 	&	$^{74}_{40}$Zr	&	HFB29	&	3.51 	&	-9.02 	&	\\
$^{34}_{20}$Ca	&	FRDM	&	0.75 	&	-3.65 	&	$^{75}_{40}$Zr	&	HFB29	&	2.51 	&	-4.63 	&	\\
$^{38}_{22}$Ti	&	FRDM	&	1.41 	&	-8.60 	&	$^{77}_{40}$Zr	&	HFB29	&	1.86 	&	-0.06 	&	\\
$^{41}_{24}$Cr	&	FRDM	&	2.12 	&	-11.10 	&	$^{78}_{42}$Mo	&	HFB29	&	3.68 	&	-8.42 	&	\\
$^{42}_{24}$Cr	&	FRDM	&	1.11	&	-4.21 	&	$^{81}_{44}$Ru	&	HFB29	&	4.78 	&	-10.70 	&	\\
$^{49}_{28}$Ni	&	FRDM	&	1.65	&	-5.80 	&	$^{82}_{44}$Ru	&	HFB29	&	3.69 	&	-7.60 	&	\\
$^{54}_{30}$Zn	&	FRDM	&	2.77	&	-10.33 	&	$^{83}_{44}$Ru	&	HFB29	&	2.33 	&	-0.88 	&	\\
$^{55}_{30}$Zn	&	FRDM	&	1.49	&	-3.05 	&	$^{84}_{44}$Ru	&	HFB29	&	2.68 	&	-3.11 	&	\\
$^{58}_{32}$Ge	&	FRDM	&	2.57	&	-9.34 	&	$^{85}_{46}$Pd	&	HFB29	&	4.47 	&	-9.11 	&	\\
$^{59}_{32}$Ge	&	FRDM	&	1.85	&	-5.36 	&	$^{86}_{46}$Pd	&	HFB29	&	3.59 	&	-6.36 	&	\\
$^{62}_{34}$Se	&	FRDM	&	2.93	&	-9.75 	&	$^{87}_{46}$Pd	&	HFB29	&	2.62 	&	-1.66 	&	\\
$^{63}_{34}$Se	&	FRDM	&	1.37	&	0.57 	&	$^{88}_{48}$Cd	&	HFB29	&	5.59 	&	-11.10 	&	\\
$^{66}_{36}$Kr	&	FRDM	&	2.65	&	-7.54 	&	$^{89}_{48}$Cd	&	HFB29	&	4.54 	&	-8.55 	&	\\
$^{70}_{38}$Sr	&	FRDM	&	3.15	&	-8.64 	&	$^{90}_{48}$Cd	&	HFB29	&	3.77 	&	-6.18 	&	\\
$^{71}_{38}$Sr	&	FRDM	&	1.83	&	-1.10 	&	$^{91}_{48}$Cd	&	HFB29	&	2.94 	&	-2.46 	&	\\
$^{74}_{40}$Zr	&	FRDM	&	3.27	&	-8.16 	&	$^{101}_{52}$Te	&	HFB29	&	5.45 	&	-9.54 	&	\\
$^{75}_{40}$Zr	&	FRDM	&	1.93	&	-0.60 	&	$^{107}_{54}$Xe	&	HFB29	&	3.08 	&	-0.24 	&	\\
$^{80}_{42}$Mo	&	FRDM	&	1.91	&	1.21 	&	$^{111}_{56}$Ba	&	HFB29	&	3.46 	&	-2.28 	&	\\
$^{83}_{44}$Ru	&	FRDM	&	2.83	&	-3.92 	&	$^{114}_{58}$Ce	&	HFB29	&	4.94 	&	-7.12 	&	\\
$^{34}_{20}$Ca	&	KTUY	&	0.96 	&	-6.53 	&	$^{116}_{58}$Ce	&	HFB29	&	3.22 	&	-0.01 	&	\\
$^{38}_{22}$Ti	&	KTUY	&	1.56 	&	-9.59 	&									

\end{tabular}
\end{ruledtabular}
\end{table*}

\begin{table*}
\caption{The competition between the true $2p$ radioactivity and $\alpha$-decay for the nuclei beyond the proton-drip line. \textit{Q}$_{\alpha }$ denotes the $\alpha$-decay energy, which is measured in MeV. The $\alpha$-decay half-lives (log$_{10}T_{1/2}^{\alpha }$) are also calculated by the GLDM.}
\centering
\begin{tabular}{cccccccc}
\hline
\hline
Nuclei & Mass Model  &  \textit{Q}$_{2p}$ (MeV) & \textit{Q}$_{\alpha }$ (MeV)  &  log$_{10}T_{1/2}^{\text{2p}}$ (s)  & log$_{10}T_{1/2}^{\alpha }$ (s) &Decay mode &\\
\hline
$^{101}_{52}$Te	&	HFB29	&	5.45 	&-0.19    &	-9.54   &	-- 	    &	$2p$	&	\\
$^{107}_{54}$Xe	&	HFB29	&	3.08 	&4.80     &	-0.24   &	-5.70 	&	$\alpha$&	\\
$^{111}_{56}$Ba	&	HFB29	&	3.46 	&4.10     &	-2.28   &	-1.40 	&	$2p$	&	\\
$^{114}_{58}$Ce	&	HFB29	&	4.94 	&4.25     & -7.12 	&	-1.11 	&	$2p$	&	\\
$^{116}_{58}$Ce	&	HFB29	&	3.22 	&4.31     &	-0.01 	&   -1.45   &   $\alpha$&    \\
\hline
\end{tabular}
\end{table*}

Within Eq. (1) and the WS4, FRDM, KTUY and HFB29 mass tables, four types of \textit{Q}$_{2p}$ values are extracted. Firstly, the experimental \textit{Q}$_{2p}$ values of $^{19}$Mg, $^{45}$Fe, $^{48}$Ni, $^{54}$Zn, and $^{67}$Kr and those from the four kinds of nuclear mass tables are listed in Table I. From Table I, we can see that
the differences between the four kinds of \textit{Q}$_{2p}$ values are very large. In addition, by comparing the experimental \textit{Q}$_{2p}$ values and those calculated ones, it is found that the experimental \textit{Q}$_{2p}$ values of $^{45}$Fe, $^{48}$Ni and $^{67}$Kr are reproduced best by the KTUY mass model. For $^{54}$Zn, the accuracy given by the HFB29 mass model is the highest. Here it is worth mentioning that the binding energies of $^{19}$Mg and its daughter nucleus ($^{17}$Ne) are not available in the WS4, FRDM, and HFB29 mass tables, the \textit{Q}$_{2p}$ value of $^{19}$Mg is estimated only by the KTUY mass table. However, the \textit{Q}$_{2p}$ value from the KTUY mass model is 0.39 MeV larger than the experimental value. Therefore, it is difficult to determine which nuclear mass model has the strongest prediction power. This indicates that there exists some uncertainty about the masses of the nuclei near the drip-line predicted by the extant nuclear mass models. It remains is a great challenge to improve the predicted abilities of the extant nuclear mass models.

Next the $2p$ radioactivity half-lives are estimated in the framework of the GLDM by inputting the four types of \textit{Q}$_{2p}$ values. In calculations, the orbital angular momenta carried by the $2p$ pair are all taken to be 0, assuming the $2p$ decay process is the fastest. To identify the true and simultaneous emission, an energy criterion was proposed in Ref.~\cite{Olsen2013}. The energy criterion of the true and simultaneous emission is \textit{Q}$_{2p}>0$ and \textit{Q}$_{p}$ $<$0.2\textit{Q}$_{2p}$.
In addition, a condition on the $2p$ decay half-lives, -7$<$ log$_{10}T_{1/2}^{\text{2p}}$ $<$-1 s (log$_{10}T_{1/2}^{\text{2p}}$ is the $2p$ decay decimal logarithm half-life.), was given to define the feasibility of experimental observation in Ref.~\cite{Olsen2013}. The lower bound of log$_{10}T_{1/2}^{\text{2p}}=10^{-7}$ s corresponds to the typical sensitivity limit of in-flight, projectile-fragmentation techniques. The upper bound of log$_{10}T_{1/2}^{\text{2p}}=10^{-1}$ s ensures that the $2p$ radioactivity will not be dominated by $\beta $ decay. However, the experimental $2p$ decay decimal logarithm half-life of the 21$^{+}$ isomer state of $^{94}$Ag is 1.90 s~\cite{Mukha2006}. Moreover, the $2p$ decay decimal logarithm half-life of $^{19}$Mg is measured as -11.40 s~\cite{Mukha2007}.
Thus, the condition on the $2p$ decay half-lives of Ref.~\cite{Olsen2013} must be extended. To avoid losing some $2p$ decay candidates, in this article an extended criterion on $2p$ decay half-lives is used, which is written as -12.00$<$log$_{10}T_{1/2}^{\text{2p}}$ $<$2.00 s. According to the energy criterion and the new criterion on $2p$ decay half-lives, the \textit{Q}$_{2p}$ values and half-lives of the true and simultaneous emission are listed in Table II.

From Table II, the probable $2p$ decay candidates can be found in the region of \textit{Z} $<$50 or \textit{Z}$\leq 50$ by each mass model.
However, the $2p$ decay modes are not be observed for the \textit{Z}=20, 26, 30 nuclides with the WS4 model and the \textit{Z}=46, 48, 50 nuclides with the FRDM model. Similarly, in the cases of the
KTUY and HFB29 mass models, the $2p$ radioactivity can not be detected by the current technique for the \textit{Z}=26, 28, 30, 44 and \textit{Z}=20, 22, 24, 26, 30 isotopes, respectively. In the region beyond \textit{Z}=50, the $2p$ decay candidates are predicted only by the HFB29 mass model, which can be seen in the last five lines of the right part of Table II. These $2p$ decay candidates are in or very close to the \textit{N}=\textit{Z} line. In addition, as can be seen from Table II, for a given nucleus the model dependence of the \textit{Q}$_{2p}$ values can also be seen evidently. Furthermore, the \textit{Q}$_{2p}$ uncertainties lead to large uncertainties of log$_{10}T_{1/2}^{\text{2p}}$ values. For example, the predicted (\textit{Q}$_{2p}$, log$_{10}T_{1/2}^{\text{2p}}$ ) values of $^{58}$Ge
are (2.63 MeV,-9.80 s), (2.57 MeV, -9.34 s), (2.45 MeV, -8.80 s), and (1.77 MeV, -4.76 s) for the WS4, FRDM, KTUY and HFB29 mass models, respectively. Finally, in Table II a general tendency on the log$_{10}T_{1/2}^{\text{2p}}$ values is found: the log$_{10}T_{1/2}^{\text{2p}}$ half-lives of the light nuclei are short and the half-lives become long for the heavy nuclei if the \textit{Q}$_{2p}$ differences of light and heavy nuclei are not so large. This is attributed to the following reason: For the light system, the Coulomb barrier between $2p$ cluster and daughter nucleus is low because of the small charge number so that the $2p$ cluster can penetrate the barrier easily. However, the Coulomb barrier becomes higher and higher with the increase of \textit{Z}. As a result, the $2p$ decay half-life is long in most cases for the heavy nuclei.

Relevant studies suggest that some neutron-deficient nuclei near \textit{N}=\textit{Z}, just above the \textit{N}=\textit{Z}=50 shell closures, exhibit
large $\alpha$-decay branches~\cite{wang2014,wyzjpg14,wyzprc17,book96}. So, it is interesting to discuss the competition between the true $2p$ radioactivity and $\alpha$-decay of the deficient-neutron nuclei beyond \textit{Z}=50.
To compare the true $2p$ decay half-lives with the $\alpha$-decay half-lives reasonably, the $\alpha$-decay half-lives are also calculated by the GLDM. In the GLDM calculations, the preformation factor of an $\alpha$-particle is used the analytic form of Ref.~\cite{zhang2009} and the angular momenta carried by the $\alpha$-particle are selected as 0. The $\alpha$-decay energies (\textit{Q}$_{\alpha }$), $\alpha$-decay half-lives (log$_{10}T_{1/2}^{\alpha }$), and competition between the two decay modes are shown in Table III. In the last column of Table III, $2p$ ($\alpha$) represents that the $2p$ radioactivity ($\alpha$-decay) is the dominant decay mode. As can be seen from Table III, $^{101}$Te, $^{111}$Ba, and $^{114}$Ce are dominated by the $2p$ radioactivity. For $^{107}$Xe and $^{116}$Ce, $\alpha$-decay is the dominant decay mode. Note that the predicted decay energies of $^{101}$Te are \textit{Q}$_{2p}$=5.45 MeV and \textit{Q}$_{\alpha }$=-0.19 MeV, so its main decay mode is the $2p$ radioactivity.

\section{Conclusions} \label{sec4}
In this article, the \textit{Q}$_{2p}$ values have been extracted by the WS4, FRDM, KTUY, and HFB29 nuclear mass models. By comparison between the four types of \textit{Q}$_{2p}$ values, it is found that the \textit{Q}$_{2p}$ values are model dependent strongly. In addition, the $2p$ radioactivity half-lives have been calculated in the framework of the GLDM by inputting different types of \textit{Q}$_{2p}$ values. As a result, the large uncertainties of
$2p$ decay half-lives are so large due to the \textit{Q}$_{2p}$ uncertainties. Moreover, according to the energy and half-life constraint conditions, the probable $2p$ decay candidates are found in the region of \textit{Z} $<$50 or \textit{Z}$\leq 50$ for all the mass models used in this article. In the region beyond \textit{Z}=50, the $2p$-decaying candidates are predicted only by the HFB29 mass model. At last, the competition between the true $2p$ radioactivity and $\alpha$-decay for the nuclei above \textit{N}=\textit{Z}=50 shell closures has been discussed. It is shown that $^{101}$Te, $^{111}$Ba and $^{114}$Ce prefer to $2p$ radioactivity and $\alpha$-decay is the dominant decay mode for $^{107}$Xe and $^{116}$Ce.

\section*{ACKNOWLEDGEMENTS}\label{sec6}

We thank professor Shangui Zhou, professor Ning Wang, and professor Fengshou Zhang for helpful discussions. This work was supported by the National Natural Science Foundation of China (Grants No. U1832120 and No. 11675265), the Natural Science Foundation for Outstanding Young Scholars of Hebei Province of China (Grants No. A2020210012 and A2018210146), the Continuous Basic Scientific Research Project (Grant No. WDJC-2019-13), and the Leading Innovation Project (Grant No. LC 192209000701).

\end{CJK}
\end{document}